# Enhancing Grid Stability with Synchronverter-Based Control for Grid-Side Converters in Wind Power Systems

Alan Valadez
*School of Electrical Engineering*
*Tel Aviv University*
Tel Aviv, Israel
alanva@tauex.tau.ac.il

George Weiss
*School of Electrical Engineering*
*Tel Aviv University*
Tel Aviv, Israel
gweiss@tauex.tau.ac.il

Chunjiang Jia
*ORE Catapult*
*Offshore House, Albert St, Blyth NE24 1LZ*
chunjiang.jia@ore.catapult.org.uk

Chong Ng
*ORE Catapult*
*Offshore House, Albert St, Blyth NE24 1LZ*
chong.ng@ore.catapult.org.uk

***Abstract*— This paper proposes a new synchronverter control strategy for wind turbine grid-side converters, aimed at bolstering grid resilience by providing essential inertial frequency support without the need for additional energy storage components. The strategy harnesses the inherent energy storage capability of the DC-link capacitor to emulate the dynamic behavior of a synchronous generator, effectively contributing to grid inertia. A dual-controller architecture is implemented, consisting of a slow PI controller to maintain DC voltage stability and a fast protection mechanism to prevent overvoltage and undervoltage excursions. Critically, the control scheme facilitates controlled reductions of up to 10% in the DC-link voltage, generating a power reserve that can be strategically deployed for inertia emulation while preserving safe operating limits. The efficacy of the proposed method is validated through simulations of a representative three-turbine wind power plant, incorporating coordinated operation via virtual friction for inter-machine damping and local-global control strategies. Results showcase the superior performance of the proposed strategy compared to conventional grid-following control, emphasizing its potential to significantly improve grid stability and address the increasing challenges posed by high renewable energy penetration.**

***Keywords— Frequency support, grid-forming control, PMSG, renewable energy integration, synchronverter, virtual inertia.***

## I. INTRODUCTION

Electrical power systems (EPS) are transitioning due to the gradual replacement of fossil-fuel-based generation with renewable energy sources such as wind power [1, 2, 3]. This transformation has introduced critical challenges, particularly reducing system inertia resulting from replacing traditional synchronous machines with inverter-based renewable generation. Reduced inertia compromises frequency stability, particularly in systems with high renewable penetration [4, 5]. Consequently, to ensure grid reliability and resilience, it is imperative to develop advanced technical solutions and regulations for integrating distributed energy resources (DERs) into the grid [6].

Traditionally, frequency control has relied on the inertial response of synchronous machines, active power regulation via governors, and load-shedding mechanisms. However, these methods become less effective in isolated grids or systems with significant renewable energy penetration [3]. Wind turbines (WTs) have emerged as potential contributors to frequency support due to their ability to store kinetic energy, offering a straightforward inertial response. Despite this, traditional approaches, such as frequency droop and virtual inertia control, face limitations in managing mechanical power effectively [3]. Moreover, restoring turbines to maximum power point tracking (MPPT) post-frequency events can induce further frequency drops, aggravating grid instability [5].

Deloading strategies, where wind turbines operate below their optimal points to reserve energy for frequency support, offer another option for providing power reserves [7, 8]. However, this strategy imposes mechanical stress, reducing turbine lifespan and efficiency. This underscores the necessity for innovative control methods that avoid deloading while enhancing system stability.

Recent advancements in power electronics have highlighted the potential of grid-forming converters to address these challenges. Synchronverter-based control, which enables converters to emulate synchronous generator dynamics, has emerged as a promising strategy. Studies such as [9, 10] have demonstrated the effectiveness of synchronverter control in improving frequency stability in doubly fed induction generator (DFIG) systems. The foundational principles of synchronverter control have been established in prior studies [11, 12], where both the rotor-side converter and the grid-side converter are operated as synchronverters. The rotor-side converter is tasked with maintaining the DC link voltage, while the grid-side converter regulates the maximum power point tracking (MPPT).

Grid-forming inverters, integrated with advanced synchronverter control strategies, further enhance system stability. For instance, [13, 14] emphasized their critical role in providing current-limiting functionalities and maintaining grid resilience. Experimental validations in [15] confirmed the effectiveness of synchronverter-based controls in addressing power quality and stability challenges.

This paper proposes a novel synchronverter-based control for permanent magnet synchronous generator (PMSG) wind turbines, focusing on the grid-side converter. The proposed control integrates virtual friction for inter-machine damping and dual-layer DC voltage regulation to emulate inertia, as outlined in [16].

The paper is structured as follows: Section II provides an overview of PMSG-based wind turbines. Section III details the proposed synchronverter-based control strategy, emphasizing DC voltage regulation. Section IV presents simulation studies and case analyses validating the proposed method. Finally, Section V concludes with insights and future research directions.

## II. Permanent Magnet Synchronous Generators (PMSG)-Based Wind Turbines

The capacity of wind turbines has significantly increased, evolving from kilowatt-scale systems to multi-megawatt installations. [17]. This steady increase in power level and grid penetration has driven the development of advanced generators, power converter systems, and control solutions to ensure seamless grid integration. Among the most prominent designs, the direct-drive permanent magnet synchronous generator (PMSG) concept has gained significant traction, primarily due to its elimination of the gearbox [7, 17]. PMSG-based wind power systems represent a pivotal trend in modern wind generation technologies [7, 17, 18].

The AC–DC–AC converter topology is widely regarded as the optimal solution for two primary objectives: maximizing power extraction from wind and delivering high-quality electricity to the grid [7, 17, 18]. Full-scale power converters in PMSG-based wind turbines are particularly advantageous, as they fully decouple the turbine from grid disturbances and provide the potential for grid support. These converters offer high reliability and superior power quality and may eliminate the need for bulky grid-side transformers, thereby reducing the weight and volume of the nacelle [17, 19].

A typical PMSG-based wind turbine features a back-to-back converter configuration, comprising a machine-side converter (MSC), a DC-link capacitor, and a grid-side converter (GSC), as illustrated in Fig. 1.

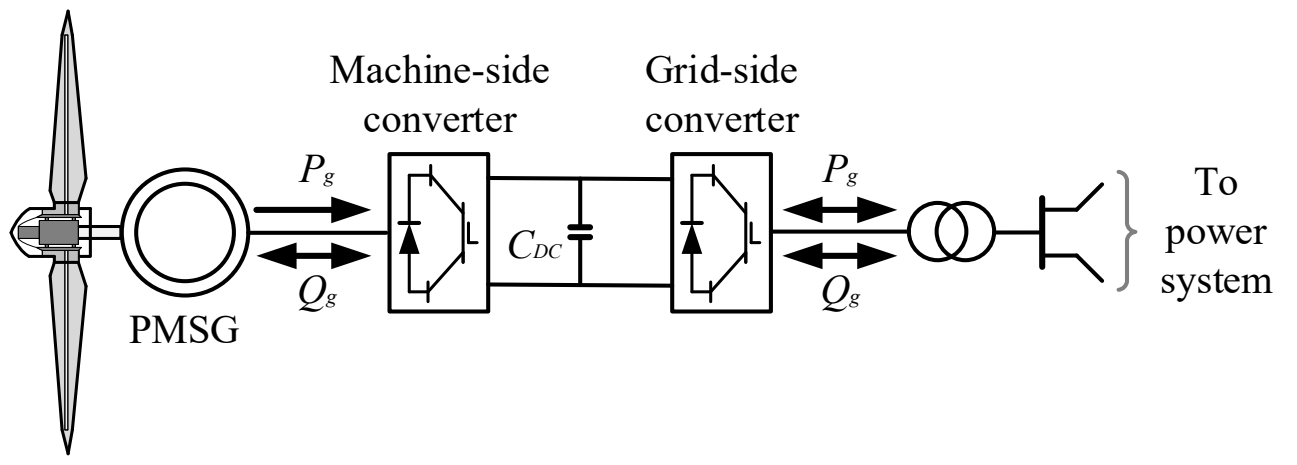

Fig. 1. A typical configuration of a direct-drive PMSG-based wind power.

### A. *Control of machine (generator)-side (MSC)/grid-side (GSC) converters*

The machine-side converter typically employs field-oriented control (FOC) for generator speed regulation. This strategy involves an outer speed control loop and an inner current control loop, often implemented using PI controllers in the synchronous reference frame [20]. The q-axis stator current component governs the electromagnetic torque, while the d-axis current is set to zero to maximize torque for a given current magnitude [21, 22].

The grid-side converter commonly uses a voltage-oriented control (VOC) strategy to regulate DC-link voltage and ensure balanced, unity power factor operation at the point of common coupling (PCC). VOC also employs dual-loop control, with an outer voltage loop and an inner current loop, enabling instantaneous decoupling of active and reactive current components in the synchronous reference frame [21]. This approach ensures fast transient response, high steady-state performance, and minimal power ripple. Traditionally, the MSC manages the power flow from the generator, while the GSC maintains the DC-link voltage and controls the power flow to the grid in compliance with grid requirements [17].

## III. Proposed Control Strategy for Synchronverter-Based Grid-Side Converter

The proposed control strategy employs a grid-forming (GFM) approach, ensuring autonomous synchronization and robustness against network disturbances [23]. This overcomes the reliance of traditional grid-following (GFL) controls on external grid conditions for synchronization.

Modern grid codes demand that power electronic devices provide frequency support mechanisms such as droop control and inertia emulation. This necessitates the adoption of GFM features in power converters. The Virtual Synchronous Machine (VSM) approach has emerged as a widely studied technique for meeting these requirements, enabling power converters to emulate the dynamics of synchronous generators. This facilitates inertia and frequency support, enhancing grid stability and enabling the seamless integration of renewable energy sources [24, 25].

### A. *Model of Synchronverter*

Fig. 2 illustrates a simplified representation of the grid-side converter of a wind turbine based on PMSG. Active and reactive power setpoints ($P_{ref}$ and $Q_{ref}$) serve as inputs to the synchronverter, which generates control signals for the pulse-width modulation (PWM) unit to operate the power converter.

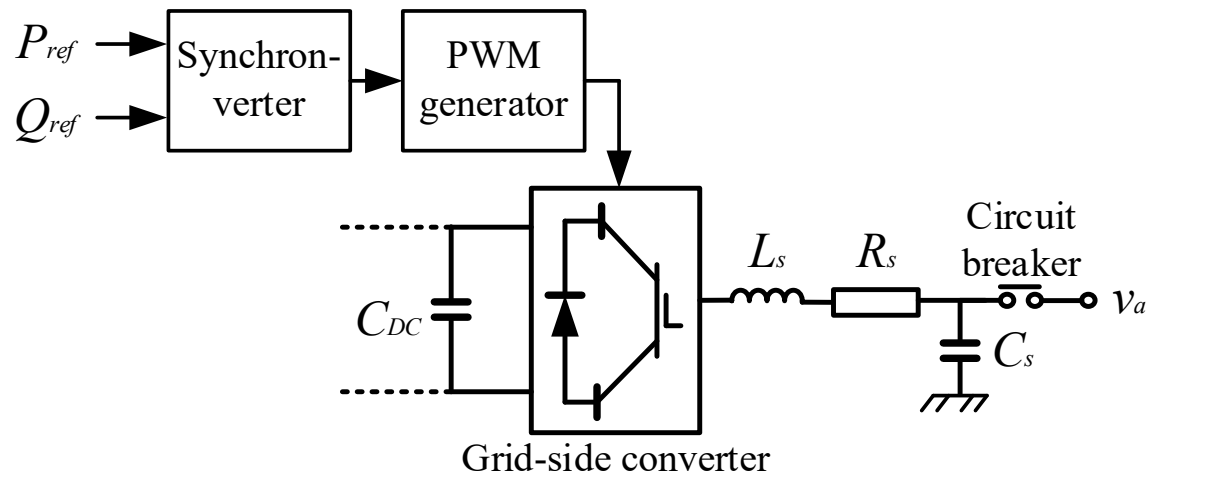

Fig. 2. The simplified power circuit of an inverter.

The synchronverter model implemented in this study is based on the framework described in [25], adapted for the grid-side converter of a wind turbine equipped with a PMSG. This model effectively replicates the dynamic behavior of a synchronous generator, allowing for frequency support using the energy stored in the DC-link capacitor, eliminating the need for dedicated energy storage hardware.

### B. *Integration of Virtual Friction*

To further enhance the stability of interconnected wind turbines, the Virtual Friction (VF) mechanism is integrated into the synchronverter model. VF is a control strategy that improves damping in VSMs by introducing an additional torque that counteracts frequency deviations. Unlike traditional damping methods (e.g., frequency droop and damper windings), VF relies on real-time communication between VSMs to calculate a global reference: the center of inertia (COI) frequency $\omega_c$.

The COI frequency is calculated as a weighted average of the frequencies of all interconnected VSMs [26]:

$$\omega_c = \frac{\sum_{k=1}^{n} \omega_j J_k}{\sum_{k=1}^{n} J_k} \tag{1}$$

Here $J_k > 0$ is the moment of inertia of VSM $k$.

By using remote frequency information, VF applies an additional damping torque to the swing equation of each VSM, proportional to the deviation of the virtual rotor frequency from $\omega_c$. This approach significantly reduces inter-machine oscillations, ensuring stable power output even during large frequency transients. Moreover, VF can emulate viscous friction in a synchronous generator, with the friction torque acting solely on the synchronverter rotor.

Fig. 3 illustrates the calculation of the damping torque $T_d$ for the VSMs. The VF torque is defined as $F\cdot(\omega-\omega_c)$, where $F$ is the VF coefficient. This damping ensures that frequency deviations among turbines are minimized, even during significant frequency events, maintaining active power output within acceptable limits and ensuring stable operation of the VSMs under dynamic conditions [27].

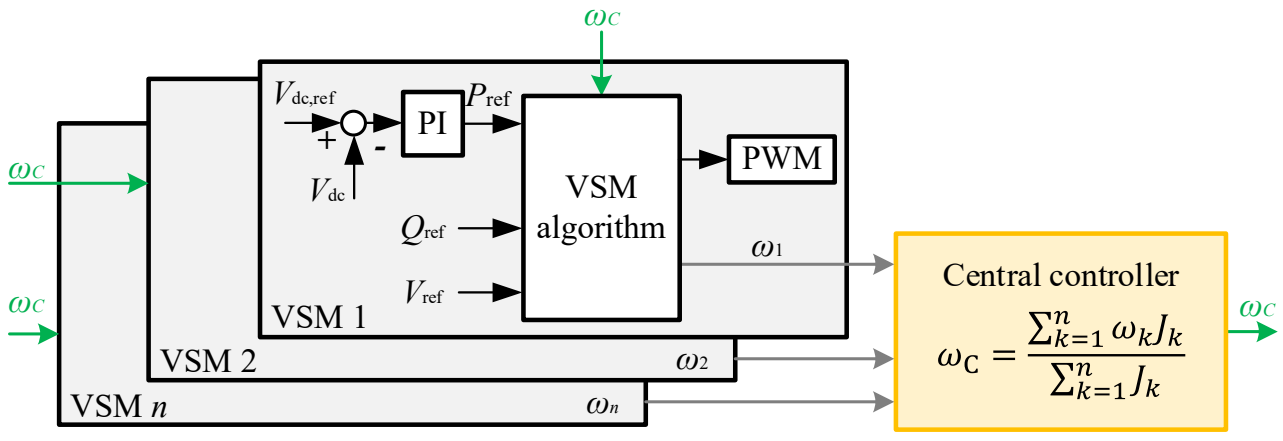

Fig. 3. Simplified block diagram of the VF in a VSM [26].

The integration of VF and COI calculation relies on real-time communication between turbines, a feature that has proven critical for managing dynamic interactions in multi-turbine configurations [26]. By combining local droop control with global VF-based damping, this strategy achieves superior performance in both isolated and grid-connected scenarios, aligning with the requirements of modern renewable energy systems.

## C. DC Voltage Regulation

The control strategy proposed for the GSC of wind turbines employs a VSM-based approach to enable frequency support without additional hardware. By leveraging the inherent flexibility of the DC-link, the system allows controlled variations in the wind turbine's total output power within a narrow range, creating an energy reserve for inertia emulation. In this study, the DC-link voltage is designed to operate within predefined limits, enabling the VSMs to emulate inertia effectively. This approach builds upon the model outlined in [16], ensuring both frequency stability and safe operation under dynamic grid conditions.

To achieve this, a dual-layer DC voltage control mechanism is implemented. The first layer consists of a slow DC voltage control loop based on a conventional proportional-integral (PI) control that regulates the DC voltage over longer timescales. This slow control ensures long-term stability by gradually restoring the voltage to its reference value after disturbances. It leverages the energy stored in DC-link capacitors for inertia emulation, adjusting the active power setpoint ($P_{ref}$) of the VSM based on the tracking error. This approach prevents abrupt changes, avoids counteracting the inertial response, and minimizes voltage excursions.

In addition, a fast voltage protection mechanism prevents DC voltage excursions beyond permissible thresholds. This mechanism directly modifies the output current controllers (OCCs) of the VSM, bypassing the slower dynamics of the PI controller. When the DC voltage drops below a lower threshold ($V_{low}$) or exceeds an upper threshold ($V_{high}$), the mechanism dynamically limits the reference current ($I_{ref}$) in the VSM to constrain the active power flow.

The protection scheme operates by adjusting the reference current $I_{ref}$ based on DC voltage conditions. The active current component ($I_a$), is calculated as the projection of the virtual current ($I_{virt}$) onto the grid voltage vector ($V$), normalized by the magnitude of the voltage:

$$I_a = \frac{V}{|V|}\cdot\frac{\langle I_{virt}, V\rangle}{|V|} \quad (2)$$

Here, the term $\langle I_{virt}, V\rangle$ represents the inner product between $I_{virt}$ and $V$. The reactive current component ($I_r$) is then obtained as the difference between the virtual current and its active current, $I_r = I_{virt} - I_a$. To address rapid DC voltage fluctuations, the mechanism incorporates voltage-dependent adjustments, defined as:

$$i_v = \frac{\langle I_{virt}, V\rangle}{|V|} \quad (3)$$

where $i_v$ can be decomposed into its positive and negative components as $i_v^+ = max\{i_v, 0\}$ and $i_v^- = max\{-i_v, 0\}$, respectively. These components are subsequently used to compute the modified reference current.

$$I_{ref} = I_a + I_r = \frac{V}{|V|} i_v^+ \beta^+ + \frac{V}{|V|} i_v^- \beta^- + I_r \quad (4)$$

Here $\beta^+$ and $\beta^-$ are dynamic limitation factors that scale the adjustments based on the DC voltage from its nominal range. As shown in Fig. 4, these factors define three distinct operating zones. In the nominal range, both $\beta^+$ and $\beta^-$ are one, and no corrective actions are required. In the transition zones, near the voltage thresholds, the factors vary linearly to apply smooth corrective adjustments. Finally, in the protection zones, the limitation factors saturate to their maximum values to strictly constrain the reference current and safeguard against over-voltage or under-voltage conditions.

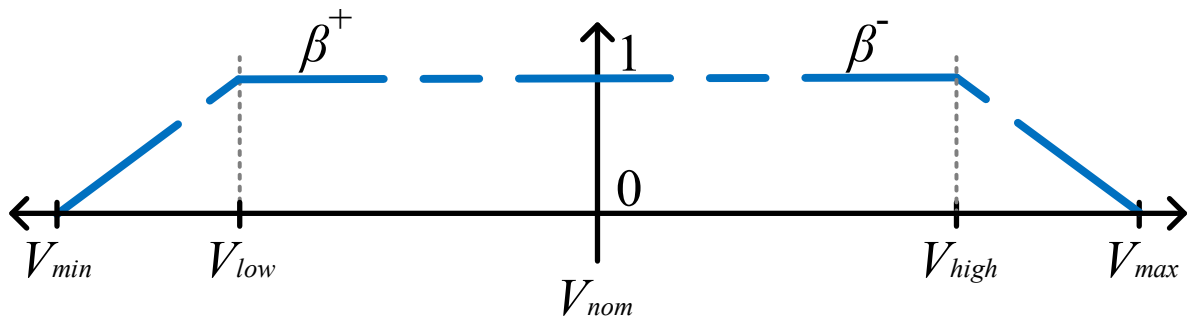

Fig. 4. The limitation factor $\beta$ as a function of the DC voltage.

Fig. 5 illustrates the phasor relationships between the grid voltage $v = v_d + jv_q$, the output current $i = i_d + ji_q$, and the internal virtual voltage of the VSM, under different power flow conditions. Fig. 5(a) shows the case where the converter injects active power, with the $dq$ reference frame aligned to the virtual rotor, highlighting how the active current component $I_a$ is modulated through the limitation factor $\beta^+$, where $\beta^+ \in [0,1]$. Conversely, Fig. 5(b) illustrates the scenario where the converter absorbs active power.

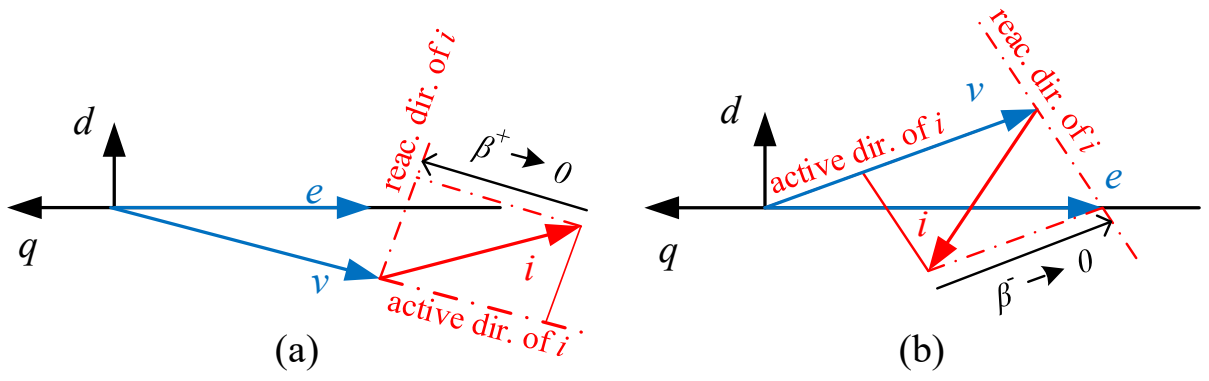

Fig. 5. Phasor diagram of the output current $i$, the grid voltage $v$ and the interval virtual voltage $e$ of the VSM: (a) when the inverter injects active power, (b) when the inverter absorbs active power.

By integrating the slow PI controller with the fast voltage protection mechanism, this dual-layer strategy achieves robust DC voltage regulation. The approach balances long-term stabilization with immediate corrective actions, enabling the system to provide effective inertia emulation and grid support while safeguarding against voltage excursions.

It is important to note that, for the wind turbine, the VSM-based control is applied exclusively to the GSC. Additionally, it is assumed that no power flows from the grid to the wind turbine, meaning the DC-link capacitor is charged solely by the energy transferred from the turbine to the grid. As a result, our control strategy focuses specifically on the inverter's behavior during power injection, as shown in Fig. 5(a), and on the influence of the limitation factor $\beta^+$ in shaping the system's dynamic response.

## IV. Case Studies

To evaluate the performance of the proposed control, we consider a simplified grid of a 15 MW synchronous generator (SG), an 8 MW aggregated load, and a 6 MW wind power plant (WPP) consisting of three 2 MW permanent magnet synchronous generator wind turbines (PMSG-WTs) as shown in Fig. 6. The synchronous generator operates at 690 V, 50 Hz, and 1800 rpm, with an inertia constant of $H$ = 1 s, and is equipped with a conventional speed governor and a constant voltage excitation system. Although the specific prime mover is not explicitly modeled, the configuration is representative of conventional thermal generation units such as gas turbines or steam turbines and thus provides a realistic benchmark for evaluating the dynamic interaction between synchronous and converter-based generation.

Each wind turbine is connected to the grid through a back-to-back converter topology, with the grid-side converter being the focus of the control strategy analysis. Two control approaches are compared: (i) a conventional grid-following (GFL) control based on a phase-locked loop (PLL) and current control loop [27], and (ii) the proposed synchronverter-based control.

The DC-link capacitor is rated at $C_{dc}$ = 200 mF, with a nominal voltage of $V_{dcnom}$ = 1200 V. The operational limits are defined as $V_{max}$ = 1.10 p.u. and $V_{min}$ = 0.90 p.u., with warning thresholds at $V_{high}$ = 1.05 p.u. and $V_{low}$ = 0.95 p.u.

The synchronverter control parameters are designed for a 2 MW base system at 690 V RMS, following the methodology in [25, 26], with a virtual inertia constant $H_v$ = 0.15 s and a virtual friction F = 5.

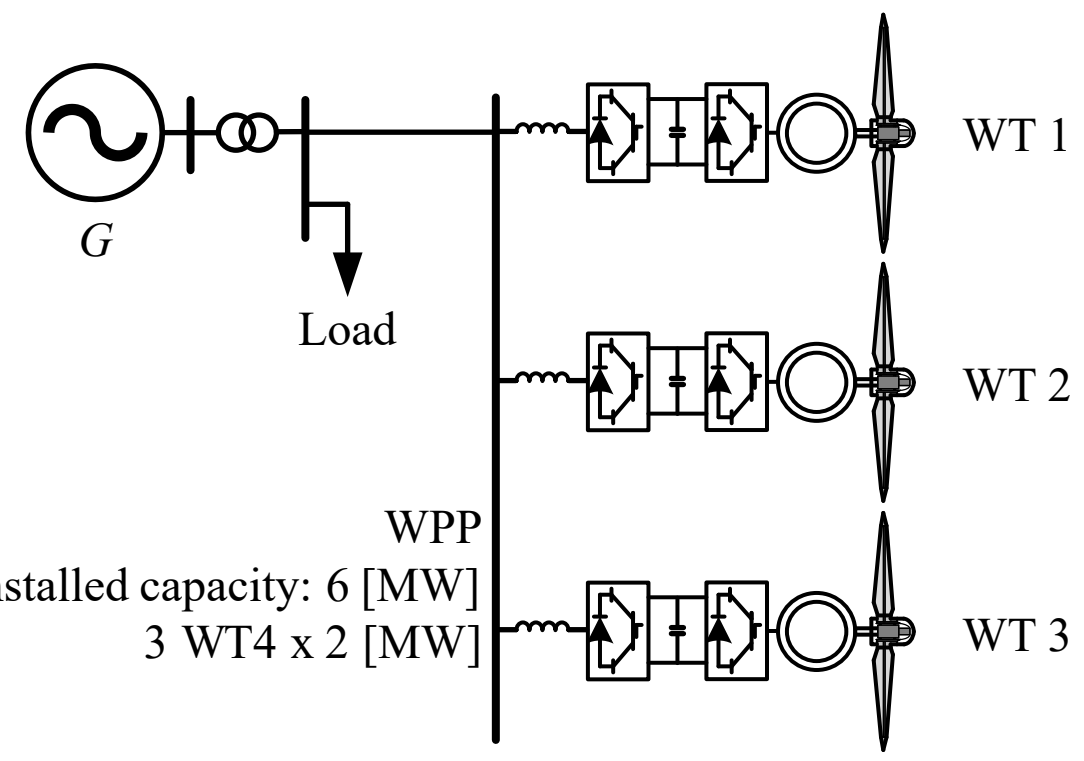

Fig. 6. Power system with integration of a wind power plan (WPP).

Simulations are conducted in MATLAB/Simulink under identical operating conditions for both control strategies. The first simulation scenario examines the system's response to a gradual increase in wind speed, from $V_w$ = 10 m/s to $V_w$ = 12 m/s, beginning at $t$ = 5 s with a 3-second linear ramp. This scenario assesses the controllers' ability to regulate the DC-link voltage under variable wind power injection—a common challenge in renewable energy systems.

Fig. 7(a) shows the active power output of a single wind turbine under both control strategies. The results indicate similar dynamic behavior across turbines, with the proposed control effectively responding to the power injection variation caused by the wind speed change. Fig. 7(b) presents the corresponding DC voltage profiles. The synchronverter-based control maintains the DC voltage within the defined boundaries, while the conventional GFL control exhibits a slower response, although it still manages to track the wind power variation.

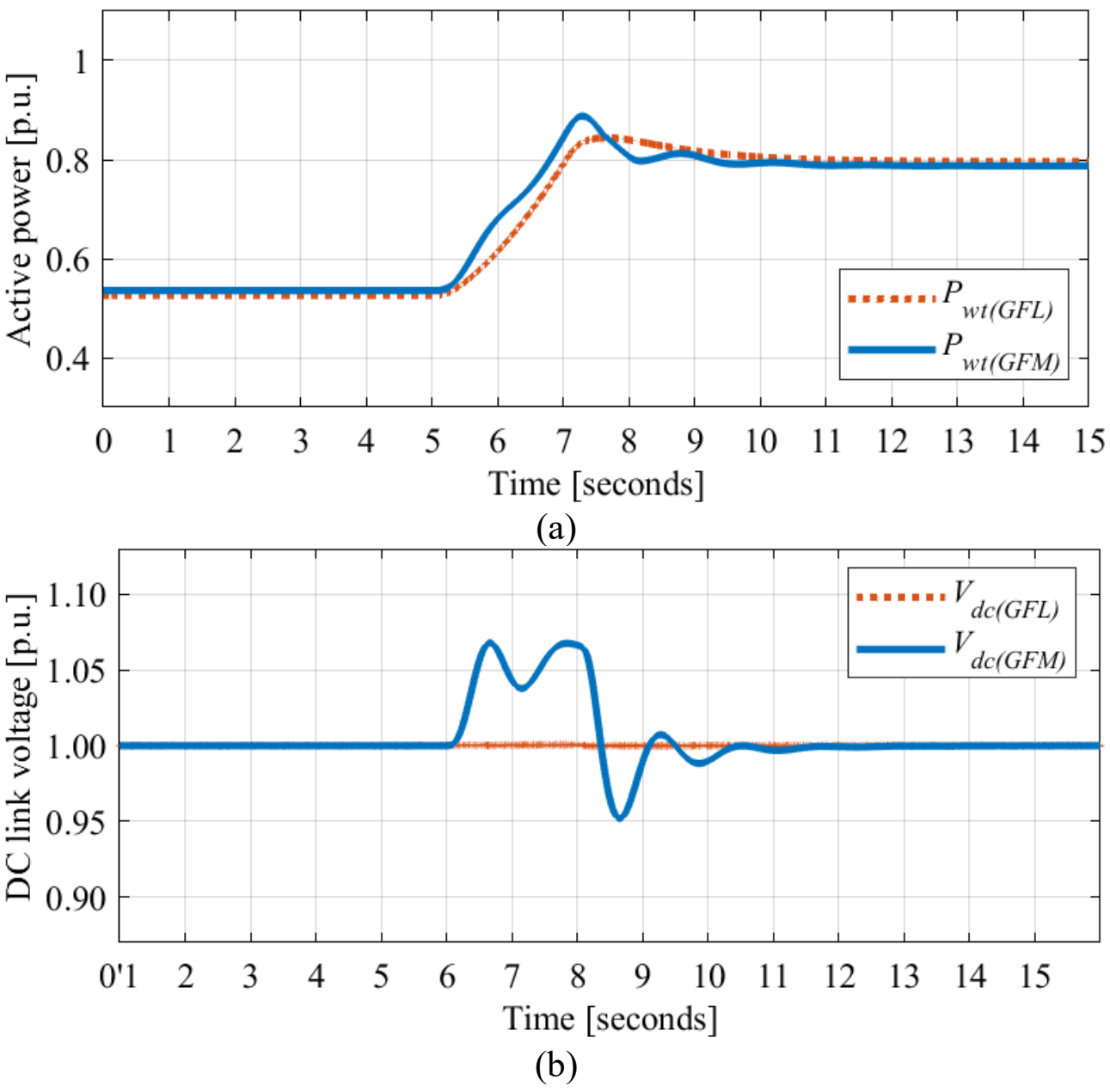

Fig. 7. (a) Active power output of a single wind turbine under different control strategies; (b) Corresponding DC-link voltage response.

To further evaluate the robustness of the proposed control under grid disturbances, a second scenario introduces a contingency event at $t$ = 5, simulating a sudden load increase that results in a frequency drop when the wind turbine is operating under a $V_w$ = 12 m/s. Fig 8(a) shows the active power of one WT under each control strategy, here we can see a rapid inertial power injection by our proposed control, reaching a peak of $P_{wt}$ = 1.048 p.u. at $t$ = 5.06 s , which helps counteract the frequency deviation.

Fig. 8(b) illustrates the DC-link voltage response during the disturbance. As the converter delivers inertial support, the voltage momentarily drops to the lower threshold of 0.9 p.u., but quickly recovers to its nominal value, demonstrating the controller's ability to manage energy exchange effectively without compromising voltage stability.

The frequency response comparison, illustrated in Fig. 9, shows that the proposed control achieves a lower frequency nadir ($f_{in}$ = 49.09 Hz at $t$ = 5.31 s) compared to the GFL control ($f_{min}$ = 49.30 Hz at $t$ = 5.28 s), indicating a more effective contribution to frequency support during transient events. This improvement is particularly relevant in low-inertia systems, where converter-based generation must increasingly take on the role of frequency regulation traditionally provided by synchronous machines.

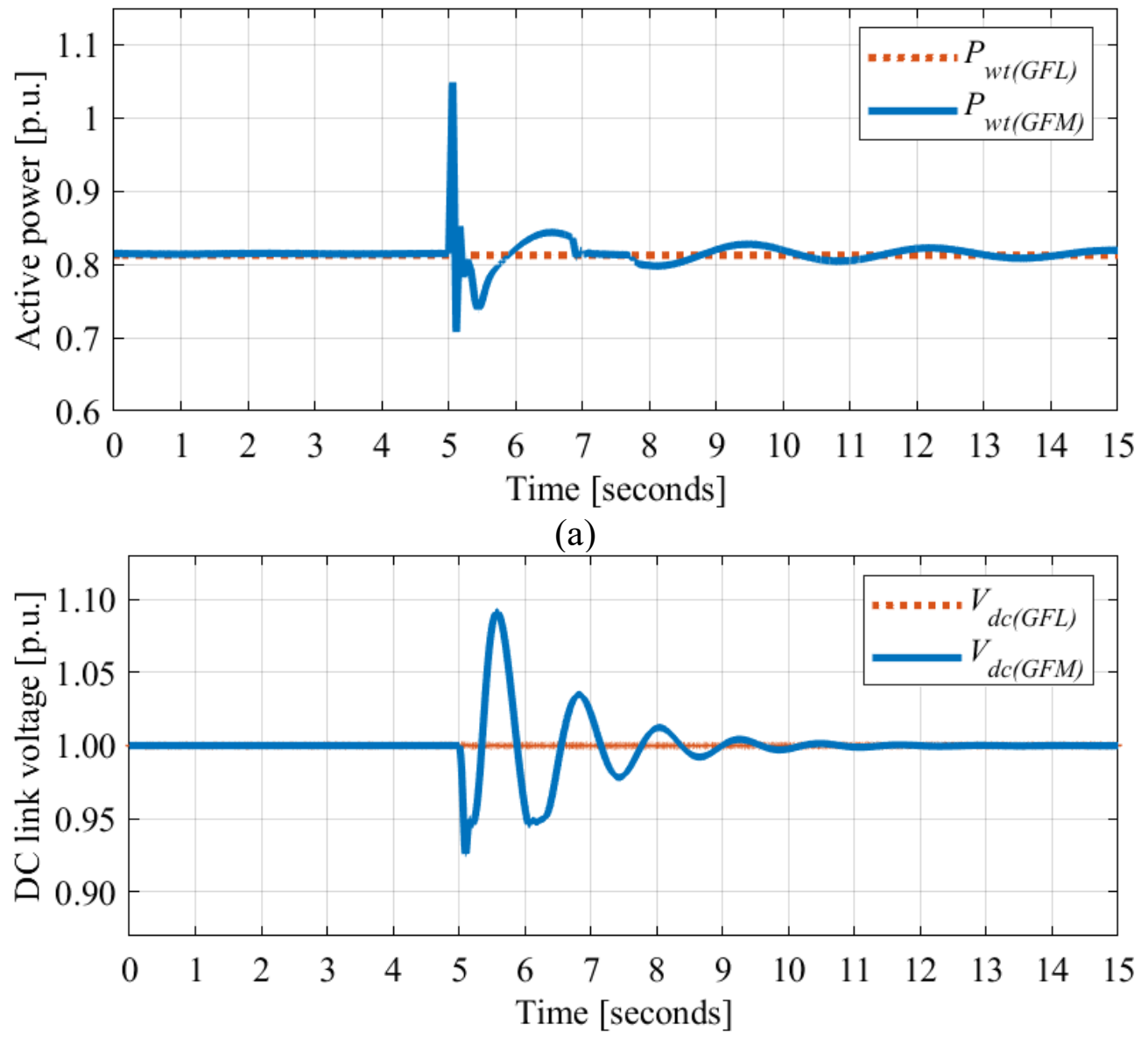

Fig. 8. (a) Active power output of DC-link voltage response under different control strategies, (b) DC link voltage response under different control strategies.

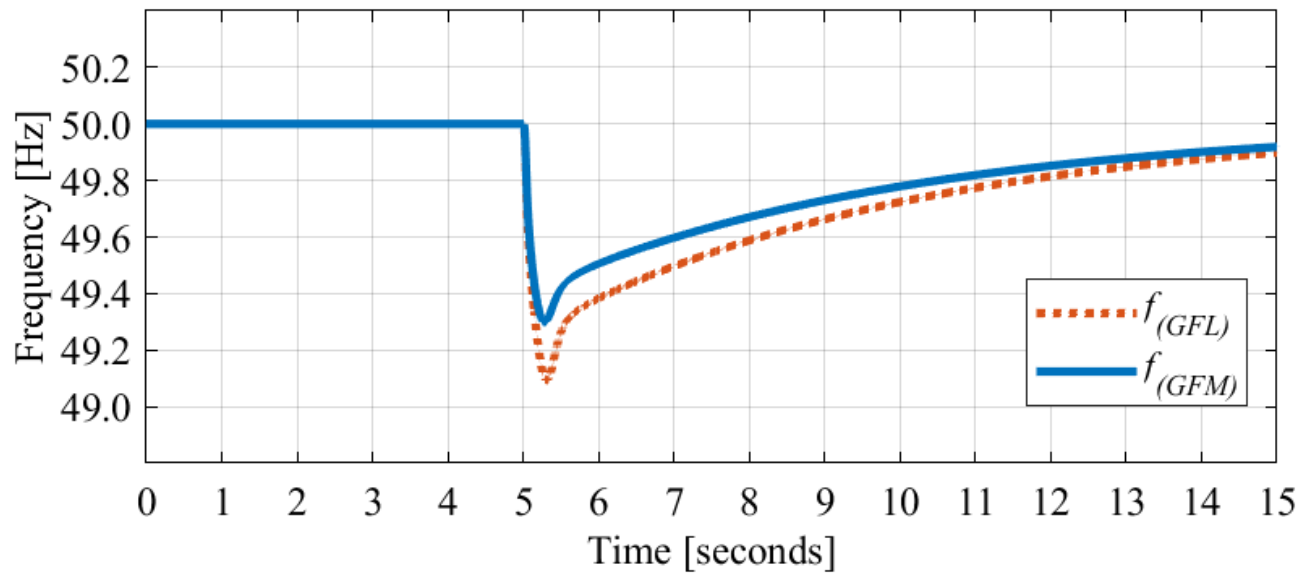

Fig. 9. Comparison of system frequency response.

Overall, the simulation results confirm that the synchronverter-based control strategy not only enhances DC-link voltage regulation under variable wind conditions but also provides effective virtual inertia support during grid disturbances. This leads to improved frequency stability and dynamic performance, making it a promising solution for integrating converter-interfaced renewable energy sources in weak or isolated power systems.

## V. Conclusions

This study has demonstrated the effectiveness of a synchronverter-based control strategy for PMSG wind turbines in enhancing grid stability and frequency support. By leveraging the energy stored in the DC-link capacitor and incorporating virtual friction mechanisms, the proposed approach enables inertia emulation and inter-machine damping, contributing to a more resilient and stable power system. Notably, this is achieved without the need for additional hardware, offering a cost-effective and practical solution for renewable energy integration. While the results are promising, validation is currently limited to a specific grid setup. Future work will extend the analysis to diverse grid conditions and include experimental testing to assess real-world feasibility and the trade-offs of DC-link voltage reduction. Further research will address scalability, integration with other renewable technologies, and comparative studies with alternative inertia emulation strategies. Additionally, coordinated control with synchronous generators will be explored to improve frequency response and load sharing in high-renewable systems.

## Acknowledgment

The work of A. Valadez is supported by the European Union's Horizon Europe research and innovation program under the Marie Sklodowska-Curie training network ModConFlex, contract number 101073558.